\pdfoutput=1
\documentclass[12pt,a4paper]{article}

\usepackage{ifthen} 
\newboolean{pdflatex}
\setboolean{pdflatex}{true} 

\newboolean{articletitles}
\setboolean{articletitles}{true} 

\newboolean{uprightparticles}
\setboolean{uprightparticles}{false} 

\def\paperauthors{LHCb collaboration} 
\def\paperasciititle{First measurement of the decay-time-integrated CP asymmetry in Bs->DsPi decays} 
\def\papertitle{First measurement of the decay-time-integrated \CP asymmetry in \BstoDspi decays} 
\def\paperkeywords{{High Energy Physics}, {LHCb}} 
\def\papercopyright{\the\year\ CERN for the benefit of the LHCb collaboration} 
\def\paperlicence{CC BY 4.0 licence}
\def\paperlicenceurl{https://creativecommons.org/licenses/by/4.0/}

\newif\ifEnableSectionTOCLinks
\EnableSectionTOCLinksfalse 

\usepackage[top=1in, bottom=1.25in, left=1in, right=1in]{geometry}

\usepackage{microtype}
\usepackage{lineno}  
\usepackage{xspace} 
\usepackage{caption} 

\usepackage{graphicx}  
\usepackage{color}
\usepackage{colortbl}
\graphicspath{{./figs/}} 

\usepackage{amsmath} 
\usepackage{amssymb}
\usepackage{amsfonts}
\usepackage{upgreek} 

\newcommand*\patchAmsMathEnvironmentForLineno[1]{%
\expandafter\let\csname old#1\expandafter\endcsname\csname #1\endcsname
\expandafter\let\csname oldend#1\expandafter\endcsname\csname
end#1\endcsname
 \renewenvironment{#1}%
   {\linenomath\csname old#1\endcsname}%
   {\csname oldend#1\endcsname\endlinenomath}%
}
\newcommand*\patchBothAmsMathEnvironmentsForLineno[1]{%
  \patchAmsMathEnvironmentForLineno{#1}%
  \patchAmsMathEnvironmentForLineno{#1*}%
}
\AtBeginDocument{%
\patchBothAmsMathEnvironmentsForLineno{equation}%
\patchBothAmsMathEnvironmentsForLineno{align}%
\patchBothAmsMathEnvironmentsForLineno{flalign}%
\patchBothAmsMathEnvironmentsForLineno{alignat}%
\patchBothAmsMathEnvironmentsForLineno{gather}%
\patchBothAmsMathEnvironmentsForLineno{multline}%
\patchBothAmsMathEnvironmentsForLineno{eqnarray}%
}

\usepackage[pdftex,
            pdfauthor={\paperauthors},
            pdftitle={\paperasciititle},
            pdfkeywords={\paperkeywords}]{hyperref}
\usepackage{hyperxmp}
\hypersetup{
    pdfcopyright={Copyright (C) \papercopyright},
    pdflicenseurl={\paperlicenceurl}
}

\usepackage[colorinlistoftodos,textsize=scriptsize]{todonotes}

\usepackage[bottom,flushmargin,hang,multiple]{footmisc}

\usepackage[all]{hypcap} 

\usepackage{xspace} 
\usepackage{upgreek}

\def\lhcb   {\mbox{LHCb}\xspace}

\def\MagUp {\mbox{\em Mag\kern -0.05em Up}\xspace}

\def\lone   {L0\xspace}

\def\hltone {HLT1\xspace}
\def\hlttwo {HLT2\xspace}

\ifthenelse{\boolean{uprightparticles}}%
{

 \def\Pmu         {\ensuremath{\upmu}\xspace}

 \def\Ppi         {\ensuremath{\uppi}\xspace}

 \def\Ppsi        {\ensuremath{\uppsi}\xspace}

 \def\PDelta      {\ensuremath{\Delta}\xspace}                 
 \def\PXi         {\ensuremath{\Xi}\xspace}                 
 \def\PLambda     {\ensuremath{\Lambda}\xspace}                 
 \def\PSigma      {\ensuremath{\Sigma}\xspace}                 
 \def\POmega      {\ensuremath{\Omega}\xspace}                 
 \def\PUpsilon    {\ensuremath{\Upsilon}\xspace}
 \let\oldPi\Pi
 \def\PPi         {\ensuremath{\oldPi}\xspace}

 \def\PB      {\ensuremath{\mathrm{B}}\xspace}                 
 \def\PD      {\ensuremath{\mathrm{D}}\xspace}                 
 \def\PJ      {\ensuremath{\mathrm{J}}\xspace}                 
 \def\PK      {\ensuremath{\mathrm{K}}\xspace}                 
 \def\Pb      {\ensuremath{\mathrm{b}}\xspace}                 
 \def\Pc      {\ensuremath{\mathrm{c}}\xspace}                 
 \def\Pd      {\ensuremath{\mathrm{d}}\xspace}

 \def\Ph      {\ensuremath{\mathrm{h}}\xspace}                 
 \def\Pp      {\ensuremath{\mathrm{p}}\xspace}                 

 \def\Ps      {\ensuremath{\mathrm{s}}\xspace}                 
                  
 \def\Pu      {\ensuremath{\mathrm{u}}\xspace}                 
 \def\thebaroffset{0.0em}
}
{

 \def\Pmu         {\ensuremath{\mu}\xspace}

 \def\Ppi         {\ensuremath{\pi}\xspace}

 \def\Ppsi        {\ensuremath{\psi}\xspace}                 
                  
 \mathchardef\PDelta="7101
 \mathchardef\PXi="7104
 \mathchardef\PLambda="7103
 \mathchardef\PSigma="7106
 \mathchardef\POmega="710A
 \mathchardef\PUpsilon="7107
 \mathchardef\PPi="7105
 \def\PB      {\ensuremath{B}\xspace}                 
 \def\PD      {\ensuremath{D}\xspace}                 
 \def\PJ      {\ensuremath{J}\xspace}                 
 \def\PK      {\ensuremath{K}\xspace}                 
 \def\Pb      {\ensuremath{b}\xspace}                 
 \def\Pc      {\ensuremath{c}\xspace}                 
 \def\Pd      {\ensuremath{d}\xspace}

 \def\Ph      {\ensuremath{h}\xspace}                 
 \def\Pp      {\ensuremath{p}\xspace}                 

 \def\Ps      {\ensuremath{s}\xspace}                 
                  
 \def\Pu      {\ensuremath{u}\xspace}                 
 \def\thebaroffset{0.18em}
}
\newcommand{\offsetoverline}[2][\thebaroffset]{\kern #1\overline{\kern -#1 #2}}%

\makeatletter
\ifcase \@ptsize \relax
  \newcommand{\miniscule}{\@setfontsize\miniscule{4}{5}}
\or
  \newcommand{\miniscule}{\@setfontsize\miniscule{5}{6}}
\or
  \newcommand{\miniscule}{\@setfontsize\miniscule{5}{6}}
\fi
\makeatother

\DeclareRobustCommand{\optbar}[1]{\shortstack{{\miniscule (\rule[.5ex]{1.25em}{.18mm})}
  \\ [-.7ex] $#1$}}

\def\mup        {{\ensuremath{\Pmu^+}}\xspace}
\def\mun        {{\ensuremath{\Pmu^-}}\xspace} 

\def\uquark    {{\ensuremath{\Pu}}\xspace}

\def\dquark    {{\ensuremath{\Pd}}\xspace}

\def\squark    {{\ensuremath{\Ps}}\xspace}

\def\cquark    {{\ensuremath{\Pc}}\xspace}

\def\bquark    {{\ensuremath{\Pb}}\xspace}

\def\hadron {{\ensuremath{\Ph}}\xspace}
\def\pion   {{\ensuremath{\Ppi}}\xspace}
\def\piz    {{\ensuremath{\pion^0}}\xspace}
\def\pip    {{\ensuremath{\pion^+}}\xspace}
\def\pim    {{\ensuremath{\pion^-}}\xspace}

\def\kaon    {{\ensuremath{\PK}}\xspace}

\def\KorKbar {\kern \thebaroffset\optbar{\kern -\thebaroffset \PK}{}\xspace}

\def\Kp      {{\ensuremath{\kaon^+}}\xspace}
\def\Km      {{\ensuremath{\kaon^-}}\xspace}

\def\KS      {{\ensuremath{\kaon^0_{\mathrm{S}}}}\xspace}

\def\D       {{\ensuremath{\PD}}\xspace}

\def\DorDbar {\kern \thebaroffset\optbar{\kern -\thebaroffset \PD}\xspace}
\def\Dz      {{\ensuremath{\D^0}}\xspace}

\def\Dp      {{\ensuremath{\D^+}}\xspace}
\def\Dm      {{\ensuremath{\D^-}}\xspace}

\def\DpDm    {\ensuremath{\Dp {\kern -0.16em \Dm}}\xspace}

\def\Dstarp  {{\ensuremath{\D^{*+}}}\xspace}

\def\Dsp     {{\ensuremath{\D^+_\squark}}\xspace}
\def\Dsm     {{\ensuremath{\D^-_\squark}}\xspace}

\def\Dssm    {{\ensuremath{\D^{*-}_\squark}}\xspace}

\def\B       {{\ensuremath{\PB}}\xspace}
\def\Bbar    {{\ensuremath{\offsetoverline{\PB}}}\xspace}

\def\BorBbar {\kern \thebaroffset\optbar{\kern -\thebaroffset \PB}\xspace}
\def\Bz      {{\ensuremath{\B^0}}\xspace}

\def\Bd      {{\ensuremath{\B^0}}\xspace}

\def\BdorBdbar {\kern \thebaroffset\optbar{\kern -\thebaroffset \Bd}\xspace}

\def\Bs      {{\ensuremath{\B^0_\squark}}\xspace}
\def\Bsb     {{\ensuremath{\Bbar{}^0_\squark}}\xspace}
\def\BsorBsbar {\kern \thebaroffset\optbar{\kern -\thebaroffset \Bs}\xspace}

\def\jpsi     {{\ensuremath{{\PJ\mskip -3mu/\mskip -2mu\Ppsi}}}\xspace}

\def\Y#1S{\ensuremath{\PUpsilon{(#1S)}}\xspace}

\def\proton      {{\ensuremath{\Pp}}\xspace}

\def\Lz          {{\ensuremath{\PLambda}}\xspace}
\def\Lbar        {{\ensuremath{\offsetoverline{\PLambda}}}\xspace}
\def\LorLbar     {\kern \thebaroffset\optbar{\kern -\thebaroffset \PLambda}\xspace}

\def\Lcbar       {{\ensuremath{\Lbar{}^-_\cquark}}\xspace}

\def\Lbbar        {{\ensuremath{\Lbar{}^0_\bquark}}\xspace}

\newcommand{\decay}[2]{\ensuremath{\mathinner{#1\!\to #2}}\xspace}

\def\to                 {\ensuremath{\rightarrow}\xspace}

\def\order   {{\ensuremath{\mathcal{O}}}\xspace}

\def\CP                {{\ensuremath{C\!P}}\xspace}

\def\AT#1     {\ensuremath{A_{\mathrm{T}}^{#1}}\xspace}           

\def\C#1      {\ensuremath{\mathcal{C}_{#1}}\xspace}                       
\def\Cp#1     {\ensuremath{\mathcal{C}_{#1}^{'}}\xspace}                    
\def\Ceff#1   {\ensuremath{\mathcal{C}_{#1}^{\mathrm{(eff)}}}\xspace}        
\def\Cpeff#1  {\ensuremath{\mathcal{C}_{#1}^{'\mathrm{(eff)}}}\xspace}       
\def\Ope#1    {\ensuremath{\mathcal{O}_{#1}}\xspace}                       
\def\Opep#1   {\ensuremath{\mathcal{O}_{#1}^{'}}\xspace}                    

\newcommand{\aunit}[1]{\ensuremath{\text{\,#1}}}       

\newcommand{\tev}{\aunit{Te\kern -0.1em V}\xspace}
\newcommand{\gev}{\aunit{Ge\kern -0.1em V}\xspace}
\newcommand{\mev}{\aunit{Me\kern -0.1em V}\xspace}
\newcommand{\kev}{\aunit{ke\kern -0.1em V}\xspace}
\newcommand{\ev}{\aunit{e\kern -0.1em V}\xspace}
 
\newcommand{\mevc}{\ensuremath{\aunit{Me\kern -0.1em V\!/}c}\xspace}
\newcommand{\gevc}{\ensuremath{\aunit{Ge\kern -0.1em V\!/}c}\xspace}
\newcommand{\mevcc}{\ensuremath{\aunit{Me\kern -0.1em V\!/}c^2}\xspace}
\newcommand{\gevcc}{\ensuremath{\aunit{Ge\kern -0.1em V\!/}c^2}\xspace}

\def\fb   {\ensuremath{\aunit{fb}}\xspace}
\def\invfb   {\ensuremath{\fb^{-1}}\xspace}

\newcommand{\stat}{\aunit{(stat)}\xspace}
\newcommand{\syst}{\aunit{(syst)}\xspace}

\def\order{{\ensuremath{\mathcal{O}}}\xspace}

\def\gsim{{~\raise.15em\hbox{$>$}\kern-.85em
          \lower.35em\hbox{$\sim$}~}\xspace}
\def\lsim{{~\raise.15em\hbox{$<$}\kern-.85em
          \lower.35em\hbox{$\sim$}~}\xspace}

\def\evtgen     {\mbox{\textsc{EvtGen}}\xspace}

\def\geant      {\mbox{\textsc{Geant4}}\xspace}

\def\photos     {\mbox{\textsc{Photos}}\xspace}

\def\pythia     {\mbox{\textsc{Pythia}}\xspace}

\def\tell1  {TELL1\xspace}
\def\ukl1   {UKL1\xspace}

\newcommand{\ie}{\mbox{\itshape i.e.}\xspace}

\newcommand{\lhcborcid}[1]{\href{https://orcid.org/#1}{\hspace*{0.1em}\raisebox{-0.45ex}{\includegraphics[width=1em]{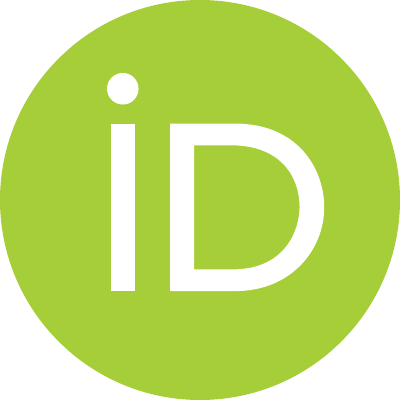}}}}

\hypersetup{
  colorlinks   = true, 
  urlcolor     = blue, 
  linkcolor    = blue, 
  citecolor    = red   
}

\ifEnableSectionTOCLinks
    \usepackage[explicit]{titlesec} 
    
    \let\oldcontentsline\contentsline
    \renewcommand

    \titleformat{\section}{\normalfont\Large\bf}{\hyperlink{tocsection.\thesection}{{\thesection} \parbox[t]{\dimexpr\textwidth-1pc}{#1}}}{1pc}{}

    \titleformat{\subsection}{\normalfont\bf}{\hyperlink{tocsubsection.\thesubsection}{{\thesubsection} \parbox[t]{\dimexpr\textwidth-1pc}{#1}}}{1pc}{}

    \titleformat{name=\section,numberless}[display]{}{}{0pt}{\normalfont\Huge\bfseries #1}
\fi

\usepackage{cite} 
\usepackage{mciteplus}

\usepackage{booktabs}
\usepackage{multirow}

\def\BstoDspi{\decay{\Bs}{\Dsm \pip}}

\def\asfs {$\mathasfs$\xspace}
\def\Asfs {{\ensuremath{A^s_{\text{fs}}}}\xspace}

\def\mathassl{{\ensuremath{a^s_{\text{sl}}}}}
\def\assl {$\mathassl$\xspace}

\def\araw{A_{\text{raw}}\xspace}

\def\BdtoDpi{\decay{\Bz}{\Dm \pip}}
\def\BdtoDspi{\decay{\Bz}{\Dsp \pim}}
\def\BstoDspi{\decay{\Bs}{\Dsm \pip}}
\def\BstoDsstpi{\decay{\Bs}{D_s^{\ast-} \pip}}

\def\BstoDsk{\decay{\Bs}{\Dsm \Kp}}

\def\LbtoLcpi{\decay{\Lbbar}{\Lcbar \pip}}

\def\BstoDspiWS{\decay{\Bs}{\Dsp\pim}}

\def\BdtoDpi {\decay{\Bz}{\Dm \pip}}

\def\bcubarq{\decay{\bquark}{\cquark \bar \uquark q}}

\def\DstoKKpi {\decay{\Dsm}{\Km\Kp\pim}}

\def\Dstopipipi {\decay{\Dsm}{\pim\pip\pim}}

\def\DtoKpipi {\decay{\Dp}{\Km\pip\pip}}
\def\DtoKspi {\decay{\Dp}{\KS\pip}}

\def\hhpiCC{\ensuremath{\hadron^-\hadron^+\pim}\xspace}

\def\hhpipiCC{\ensuremath{\hadron^-\hadron^+\pim\pip}\xspace}

\def\KKpipiCC{\ensuremath{\Km\Kp\pim\pip}\xspace}

\def\pipipipiCC{\ensuremath{\pim\pip\pim\pip}\xspace}

\def\Asfs {{\ensuremath{A^s_{\rm fs}}}\xspace}
\def\asfs {{\ensuremath{a^s_{\rm fs}}}\xspace}

\def\asuntagged {{\ensuremath{\langle A_{\text{untagged}}^{s}\rangle}}\xspace}

\def\asuntaggedKKpipiCC {{\ensuremath{\asuntagged \left(\KKpipiCC\right)}}\xspace}

\def\asuntaggedpipipipiCC {{\ensuremath{\asuntagged \left(\pipipipiCC\right)}}\xspace}
\def\asuntaggeddspi {{\ensuremath{\asuntagged \left(\Dsm\pip\right)}}\xspace}

\usepackage{booktabs}
\usepackage{multirow}

\def\BstoDspi{\decay{\Bs}{\Dsm \pip}}

\def\asfs {$\mathasfs$\xspace}
\def\Asfs {{\ensuremath{A^s_{\text{fs}}}}\xspace}

\def\mathassl{{\ensuremath{a^s_{\text{sl}}}}}
\def\assl {$\mathassl$\xspace}

\def\araw{A_{\text{raw}}\xspace}

\def\BdtoDpi{\decay{\Bz}{\Dm \pip}}
\def\BdtoDspi{\decay{\Bz}{\Dsp \pim}}
\def\BstoDspi{\decay{\Bs}{\Dsm \pip}}
\def\BstoDsstpi{\decay{\Bs}{D_s^{\ast-} \pip}}

\def\BstoDsk{\decay{\Bs}{\Dsm \Kp}}

\def\LbtoLcpi{\decay{\Lbbar}{\Lcbar \pip}}

\def\BstoDspiWS{\decay{\Bs}{\Dsp\pim}}

\def\BdtoDpi {\decay{\Bz}{\Dm \pip}}

\def\bcubarq{\decay{\bquark}{\cquark \bar \uquark q}}

\def\DstoKKpi {\decay{\Dsm}{\Km\Kp\pim}}

\def\Dstopipipi {\decay{\Dsm}{\pim\pip\pim}}

\def\DtoKpipi {\decay{\Dp}{\Km\pip\pip}}
\def\DtoKspi {\decay{\Dp}{\KS\pip}}

\def\hhpiCC{\ensuremath{\hadron^-\hadron^+\pim}\xspace}

\def\hhpipiCC{\ensuremath{\hadron^-\hadron^+\pim\pip}\xspace}

\def\KKpipiCC{\ensuremath{\Km\Kp\pim\pip}\xspace}

\def\pipipipiCC{\ensuremath{\pim\pip\pim\pip}\xspace}

\def\Asfs {{\ensuremath{A^s_{\rm fs}}}\xspace}
\def\asfs {{\ensuremath{a^s_{\rm fs}}}\xspace}

\def\asuntagged {{\ensuremath{\langle A_{\text{untagged}}^{s}\rangle}}\xspace}

\def\asuntaggedKKpipiCC {{\ensuremath{\asuntagged \left(\KKpipiCC\right)}}\xspace}

\def\asuntaggedpipipipiCC {{\ensuremath{\asuntagged \left(\pipipipiCC\right)}}\xspace}
\def\asuntaggeddspi {{\ensuremath{\asuntagged \left(\Dsm\pip\right)}}\xspace}

\usepackage{longtable} 

\begin{document}

\renewcommand{\thefootnote}{\fnsymbol{footnote}}
\setcounter{footnote}{1}


\begin{titlepage}
\pagenumbering{roman}

\vspace*{-1.5cm}
\centerline{\large EUROPEAN ORGANIZATION FOR NUCLEAR RESEARCH (CERN)}
\vspace*{1.5cm}
\noindent
\begin{tabular*}{\linewidth}{lc@{\extracolsep{\fill}}r@{\extracolsep{0pt}}}
\ifthenelse{\boolean{pdflatex}}
{\vspace*{-1.5cm}\mbox{\!\!\!\includegraphics[width=.14\textwidth]{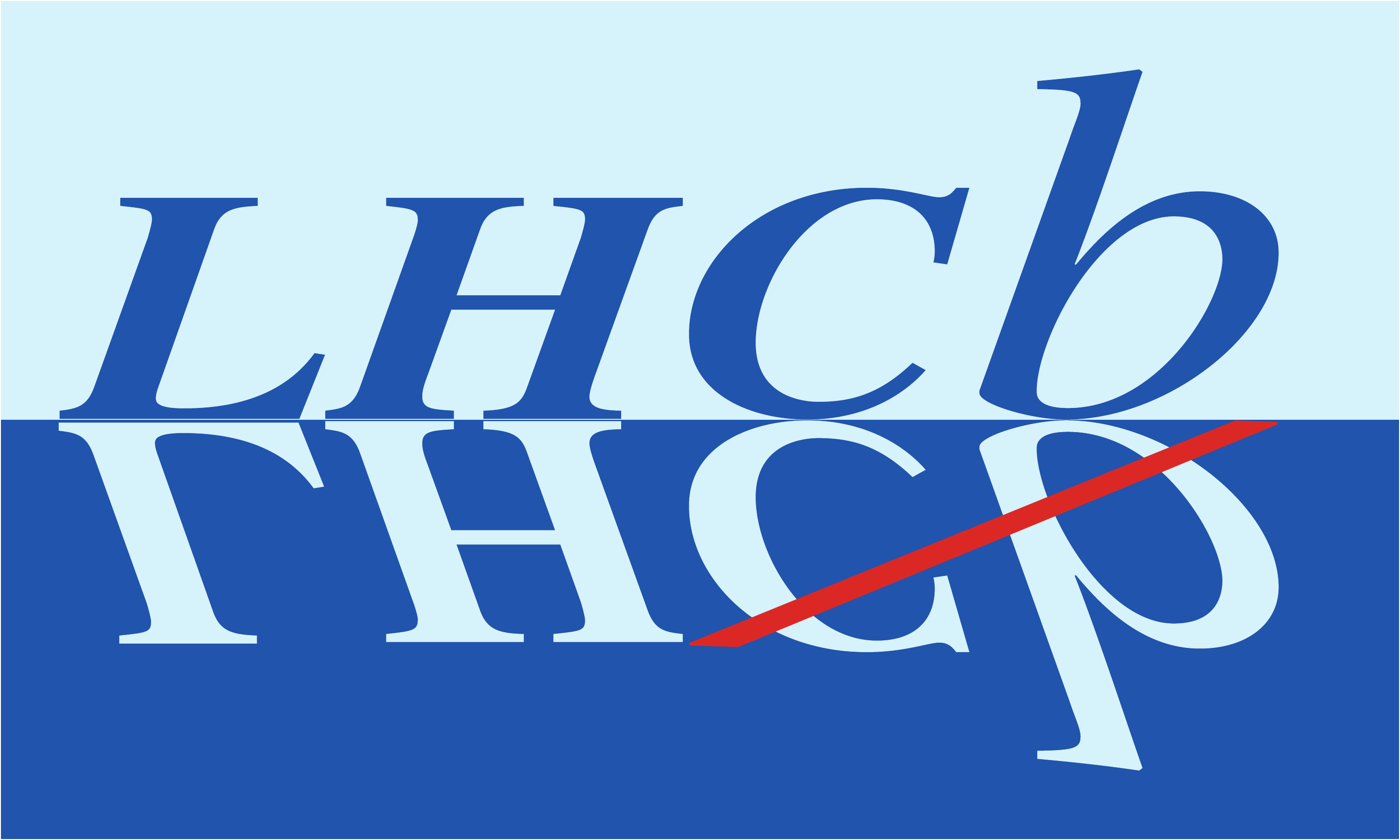}} & &}%
{\vspace*{-1.2cm}\mbox{\!\!\!\includegraphics[width=.12\textwidth]{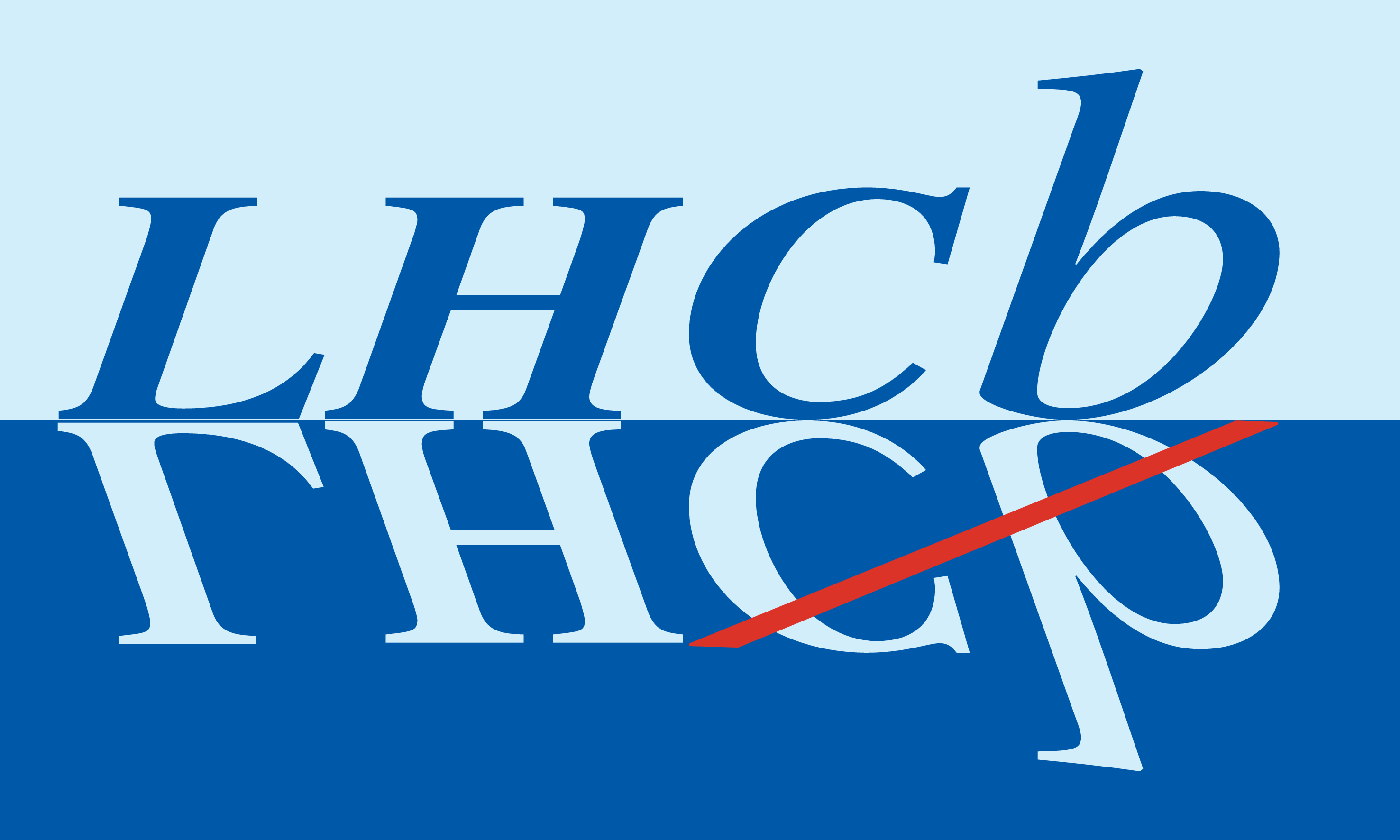}} & &}%
\\
 & & CERN-EP-2026-028 \\  
 & & LHCb-PAPER-2025-074 \\  
 & & August 25, 2026 \\ 
 & & \\
\end{tabular*}

\vspace*{4.0cm}

{\normalfont\bfseries\boldmath\huge
\begin{center}
  \papertitle 
\end{center}
}

\vspace*{2.0cm}

\begin{center}
\paperauthors\footnote{Authors are listed at the end of this Letter.}
\end{center}

\vspace{\fill}

\begin{abstract}
  \noindent
    A measurement of the flavor-untagged decay-time-integrated \CP asymmetry in the flavor-specific decay \BstoDspi, \asuntagged, is performed 
    using proton-proton collision data collected by the \lhcb experiment
    between 2016 and 2018 at a center-of-mass energy of 13\tev, corresponding to a total integrated luminosity of 5.4\invfb. 
    The \CP asymmetry is measured in two \Dsm meson decay modes, \DstoKKpi and \Dstopipipi. The combined result, \mbox{\asuntagged $= ( -1.4 \pm 5.9\stat \pm 1.1\syst) \times 10^{-3}$}, is consistent with the Standard Model expectation and provides a direct constraint on new physics in tree-level \bquark-hadron decays.

\end{abstract}

\vspace*{2.0cm}

\begin{center}
  Published in Phys.~Rev.~Lett. 137 081802 (2026)
\end{center}

\vspace{\fill}

{\footnotesize 
\centerline{\copyright~\papercopyright. \href{\paperlicenceurl}{\paperlicence}.}}
\vspace*{2mm}

\end{titlepage}


\newpage
\setcounter{page}{2}
\mbox{~}
%
%
%
%


\renewcommand{\thefootnote}{\arabic{footnote}}
\setcounter{footnote}{0}


\cleardoublepage


\pagestyle{plain} 
\setcounter{page}{1}
\pagenumbering{arabic}


Decays of beauty mesons to open-charm mesons and an associated light hadron of the form $B^0_{(s)}\to D_{(s)}^{(*)-}h^+$, where $h\in\{\pi, K\}$, are dominated by color-allowed, tree-level amplitudes proceeding through the \bcubarq quark transition, for $q\in \{\dquark, \squark\}$.\footnote{Charge-conjugate modes are implied throughout, unless otherwise stated.}
Theoretical predictions of their branching fractions in the Standard Model (SM), based on quantum chromodynamics factorisation, consistently disagree with experimental averages~\cite{Iguro2020, Bordone2020, meiser-vandyk-virto-btocubarq, Fleischer:2021cwb, Fleischer:2021cct, Fleischer:2010ca, Cai:2021mlt}.
One possible explanation is physics beyond the SM (BSM), which may introduce an additional weak phase into the decay amplitude~\cite{Skidmore2022, Huber:2016xod, Fleischer:2016dqd, Dunietz:2000cr}.
The interference of such a BSM amplitude with the SM amplitude can induce \CP violation, dubbed direct \CP violation, resulting in a small decay-rate asymmetry.

As the decay \BstoDspi is flavor specific (the corresponding wrong-sign decay \BstoDspiWS is heavily suppressed) and proceeds only at tree level, direct \CP violation is not possible in the SM as this would require interference with additional amplitudes.
The flavor-specific \CP asymmetry, \ie, the asymmetry of measuring \CP-conjugate final states $f$ and $\bar{f}$, is defined as a function of time, $t$, as
\begin{equation}
    \Asfs(t) \equiv \frac{
        \Gamma\!\left(\Bsb(t) \to \overline{f}\right) - \Gamma\!\left(\Bs(t) \to f\right)
    }{
        \Gamma\!\left(\Bsb(t) \to \overline{f}\right) + \Gamma\!\left(\Bs(t)\to f\right)
},
\end{equation}
wherein $\Gamma$ is the partial decay rate of a given state.
This is equal to the \CP asymmetry in \Bs--\Bsb mixing in the SM, denoted as \asfs for flavor-specific nonleptonic decays.
This quantity has been measured in semileptonic decays, for which it is denoted by \assl, as \mbox{$\assl = \left(-60 \pm 280\right) \times 10^{-5}$} \cite{HFLAV23}, consistent with the SM prediction of \mbox{$\assl = \left(2.06 \pm 0.18\right)\times 10^{-5}$}~\cite{Lenz2020}.

The \CP asymmetry of \BstoDspi decays can be measured as a decay-time-integrated flavor-untagged quantity, \asuntagged, to make use of the fast oscillation frequency of the \Bs meson and to circumvent the low efficiency to tag the initial \BsorBsbar flavor~\cite{LHCb-PAPER-2021-005}.
This is evaluated as
\begin{equation}
    \asuntagged \approx A^s_{\rm dir} - \frac{\asfs}{2}(1 - \rho_s),
    \label{eqn:As_untagged_mixing}
\end{equation}
wherein $A^s_{\rm dir}$ is a direct \CP-asymmetry contribution from potential BSM effects, which can be up to $\order(10^{-2})$\cite{Skidmore2022}, and $\rho_s \approx 0.001$ accounts for dilution from \Bs--\Bsb oscillation \cite{Skidmore2022}.

Experimentally, \asuntagged can be measured from the raw asymmetry, $\araw$, of observed \BstoDspi candidate yields, $N$, to final states $\Dsp\pim$ and $\Dsm\pip$, obtained through fits to the decay-time-integrated invariant-mass distributions,
\begin{equation}
        \araw ~=~ 
        \frac{N(\Dsp\pim) - N(\Dsm\pip)}
        {N(\Dsp\pim) + N(\Dsm\pip)}
        \label{eqn:A_raw}.
\end{equation}
This asymmetry is subject to charge-dependent differences in detection efficiency, resulting in a detection asymmetry, $A_{\rm det}$, which must be subtracted as
\begin{equation}
    \langle A_{\text{untagged}}^{s}\rangle = A_{\rm raw} - A_{\rm det}.
    \label{eqn:As_untagged}
\end{equation}
The term $A_{\rm det}$ is determined from data using calibration and simulated samples. 
In this measurement, the decay-time-integrated \Bs production asymmetry is neglected, as this is diluted to $\order(10^{-5})$  by the fast \Bs--\Bsb oscillation frequency \cite{LHCb-PAPER-2011-029, LHCb-PAPER-2013-033, LHCb-PAPER-2021-005}.

This Letter presents a measurement of the flavor-untagged decay-time-integrated \CP asymmetry of \BstoDspi decays, \asuntagged, performed using proton-proton $(\proton\proton)$ collision data collected by the \lhcb experiment between 2016 and 2018 at a center-of-mass energy of 13\tev, corresponding to a total integrated luminosity of 5.4\invfb.

The LHCb detector~\cite{LHCb-DP-2008-001,LHCb-DP-2014-002} is a single-arm forward spectrometer covering the pseudorapidity range $2 < \eta < 5$, designed for the study of particles containing \bquark\ or \cquark\ quarks.
The detector elements of particular relevance to this analysis include: a silicon-strip vertex detector surrounding the $pp$ interaction region that allows \cquark\ and \bquark\ hadrons to be identified from their characteristically long flight distance; a tracking system that provides a measurement of the momentum, $p$, of charged particles from their deflection by a large dipole magnet (the polarity of which is periodically reversed to reduce detection asymmetries); and two ring-imaging Cherenkov detectors that can discriminate between different species of charged hadrons.

The online event selection is performed by a trigger~\cite{LHCb-DP-2019-001}, 
which consists of a hardware stage, \lone, based on information from the electromagnetic and hadronic (HCAL) calorimeter and muon systems, followed by two software stages, \hltone and \hlttwo, which perform partial and full event reconstruction, respectively.
At the \lone stage, events must contain a final-state hadron with greater transverse energy than a threshold varying between 3.46 and 3.70\gev depending on the data-taking year according to the detector occupancy.
A minimal spatial separation between final-state hadrons in the HCAL is required, such that energy deposits do not overlap.
At the \hltone stage, single or pairs of tracks are required to be significantly displaced from any primary $pp$ interaction vertex (PV), as is indicative of a \bquark\ hadron decay.
Similarly, \hlttwo requires a two-, three- or four-track secondary vertex (SV) with a significant displacement from any PV.
A multivariate algorithm~\cite{BBDT,LHCb-PROC-2015-018} is used to select those SVs consistent with the decay topology of a \bquark hadron.
Triggered events undergo a centralized, offline processing step and highly automated \lhcb analysis productions deliver physics-analysis-ready data across the entire \lhcb physics program~\cite{Stripping, Sprucing}.

Simulated samples are used to determine reconstruction and selection efficiencies and asymmetries, and shapes of invariant-mass distributions.
In the simulation, $pp$ collisions are generated using \pythia~\cite{Sjostrand:2007gs,*Sjostrand:2006za} with a specific \lhcb configuration~\cite{LHCb-PROC-2010-056}.
Decays of unstable particles are described by \evtgen~\cite{Lange:2001uf}, in which final-state radiation is generated using \photos~\cite{davidson2015photos}.
The interaction of the generated particles with the detector, and its response, are implemented using the \geant toolkit~\cite{Allison:2006ve, *Agostinelli:2002hh} as described in Ref.~\cite{LHCb-PROC-2011-006}. 
These samples are corrected to account for known differences between data and simulation in the \Bs production kinematics and detector occupancy, and track-reconstruction, particle-identification and trigger efficiencies.

The \BstoDspi decay is reconstructed by selecting \Dsm candidates through their decays to $\Km\Kp\pim$ and $\pim\pip\pim$ final states, and combining them with a charged pion, referred to as the companion particle.
Both final states are charge symmetric under the exchange of kaon or pion pairs.
The invariant mass of all final-state particles, $m(\hhpipiCC)$, is required to be within the range $5280\text{--}6000\mevcc$ and the $\hhpiCC$ invariant mass must lie within 30\mevcc of the known \Dsm mass~\cite{PDG2024}. 
To separate the \BstoDspi signal from background sources with misidentified particles, particle identification (PID) requirements are applied using multivariate techniques that combine PID and tracking information from all \lhcb subdetectors into two discriminants per particle species~\cite{LHCb-DP-2018-001}.
A loose PID requirement is applied on the companion track, since the \BstoDspi decay has a branching fraction roughly an order of magnitude larger than the misidentified \BstoDsk decay.
The PID requirements on the \Dsm final-state tracks are optimized separately for the $\DstoKKpi$ and $\Dstopipipi$ decay modes.
The $\Km\Kp\pim$ phase space is divided into three regions: \decay{\Dsm}{ \phi\pim}, \decay{\Dsm}{K^{*}(892)^{0} \Km}, and nonresonant, while the $\pim\pip\pim$ phase space is split into \decay{\Dsm}{f_0(980)\pim} and nonresonant regions.
Loose PID requirements are applied in resonant regions, where background contamination is low, while tight criteria are set in the nonresonant regions.
In addition, dedicated vetoes are applied to suppress candidates from background contributions containing $\jpsi$ and $\Dz$ particles.
Residual background components are modeled in the invariant-mass fit.
After the selection, approximately 0.5\% and 2.5\% of events contain multiple candidates in the $\DstoKKpi$ and $\Dstopipipi$ final states, respectively.
In these events, one candidate is retained arbitrarily, preferentially choosing candidates reconstructed in the $\DstoKKpi$ final state where events remain reconstructed in both final states.
This selection strategy differs from that of Ref.~\cite{LHCb-PAPER-2021-005} in the specific PID requirements applied, in the requirement of hadron-pair separation in the HCAL, and in that a multivariate classifier was not used to suppress combinatorial background.

The total \BstoDspi yield and $\araw$ are determined from unbinned extended maximum-likelihood fits to $m(\hhpipiCC)$ across both spectra, and $A_{\rm raw}$ is determined through a simultaneous fit to the $\Dsp\pim$ and $\Dsm\pip$ invariant-mass distributions, where the models describing each component are shared across the invariant-mass distributions.
Signal and background probability density functions (PDFs) for the invariant-mass fit are obtained from dedicated simulated and data samples, with the simulated samples corrected for the \lone and PID responses to better reflect the distributions observed in data
The \lone responses are calibrated using dedicated control samples of $\Dstarp\to\Dz\pip$ decays; the PID responses in simulation are calibrated using dedicated control samples of $\Dstarp\to\Dz\pip$ and $\Lz\to \proton\pim$ decays~\cite{LHCb-DP-2018-001}. 

As in Ref.~\cite{LHCb-PAPER-2021-005}, the $m(\hhpipiCC)$ distribution for signal candidates is modeled as the sum of a modified Hypatia function with tails on both sides of the peak~\cite{Santos:2013gra} and a Johnson $S_U$ function~\cite{johnson1949systems}, sharing a common peak position. 
Signal PDFs are determined from simulated \BstoDspi decays for each \Dsm final state separately.
The PDF parameters are fixed in the fits to data, except for additive contributions to the peak position and width parameters, which account for resolution differences between data and simulation and are allowed to vary in the fit.
The total \BstoDspi yield and $\araw$ are also allowed to vary in the fit. 

The combinatorial background consists of random track combinations.
It is modeled using two exponential functions derived from fits to the upper $m(\hhpipiCC)$ sideband region, $5600\text{--}6800\mevcc$.
In the baseline fit, all shape parameters are fixed except for the larger exponent, which is left free to account for possible differences between the upper sideband and the baseline fit region.
The yield and asymmetry of the combinatorial background are determined directly from the data.

The background description closely follows Ref.~\cite{LHCb-PAPER-2021-005} and includes fully reconstructed \BdtoDspi decays, \BstoDsk decays in which the kaon is misidentified as a pion, and partially reconstructed $\BstoDsstpi$ decays where the neutral particle from ${\Dssm\to \Dsm \gamma/\piz}$ decays is not reconstructed.
Because of the similarity in the parametrization of \BdtoDspi and \BstoDsstpi decays in the selected mass range, the two decays are represented by the same distribution, \ie taking the \BdtoDspi shape as that of \BstoDspi decays and adjusting for the known mass difference between the \Bd and \Bs mesons~\cite{PDG2024}.
Additional background contributions in the \DstoKKpi mode arise from \BdtoDpi and \LbtoLcpi decays, where misidentification leads to the charm hadron being reconstructed as a \Dsm meson.
The PDFs for these components are determined from simulation, with only the peak position and width parameters varying in the baseline fit, using the same additive shifts as the signal description.
In the baseline fit, misidentified background asymmetries are fixed to zero to improve the stability of the fit, while the asymmetry of the component describing \BdtoDspi and \BstoDsstpi decays, which contains many more events, is free to vary.
The yields of \BstoDsk and \BdtoDpi decays are fixed relative to the signal yields, corrected with their corresponding efficiencies from simulation and known branching fractions.

\begin{figure}[!tb]
  \centering
  \includegraphics[width=.92\textwidth, clip, trim={0 0.125cm 1.5cm 0.125cm}]{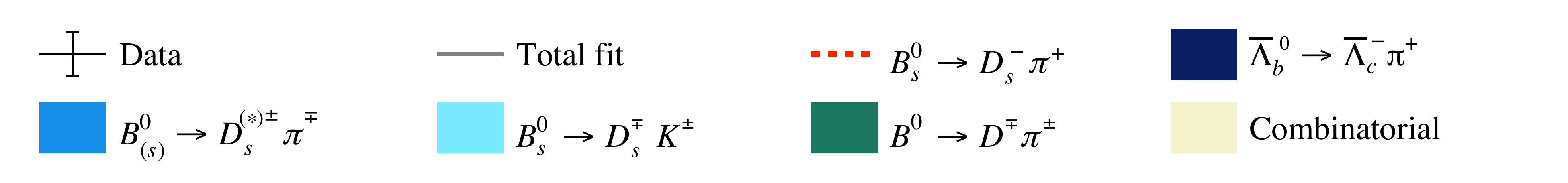}
  \includegraphics[width=.465\textwidth, clip, trim={0 0.25cm 0 0.425cm}]{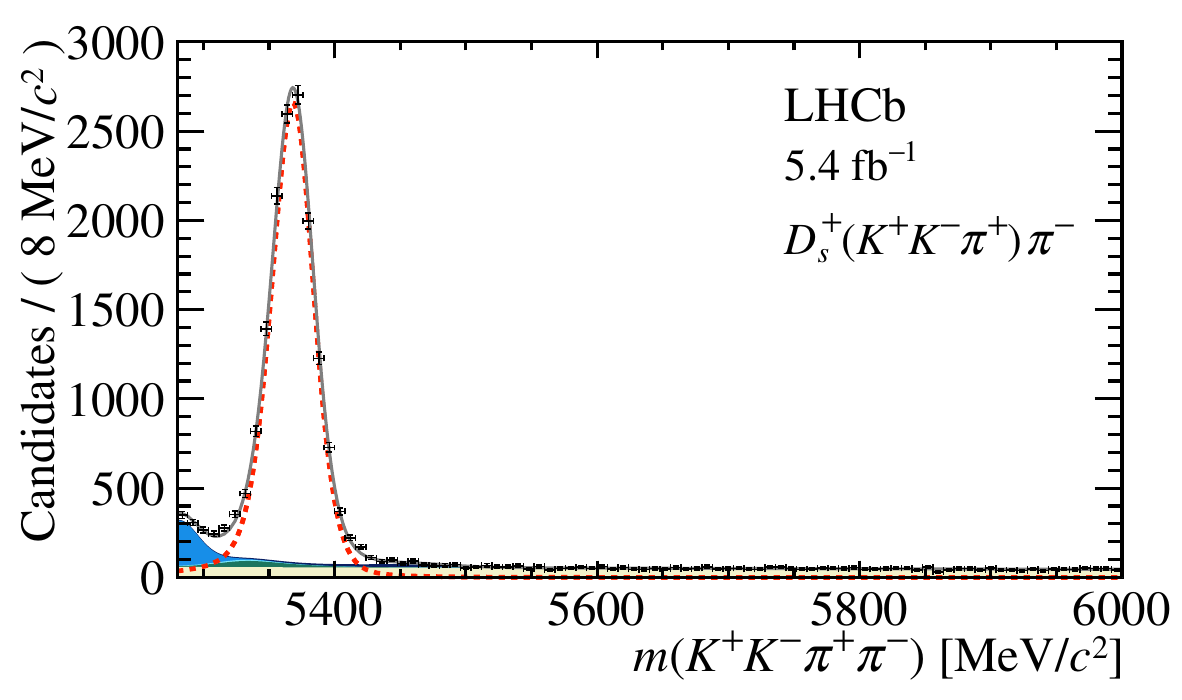}
  \includegraphics[width=.465\textwidth, clip, trim={0 0.25cm 0 0.425cm}]{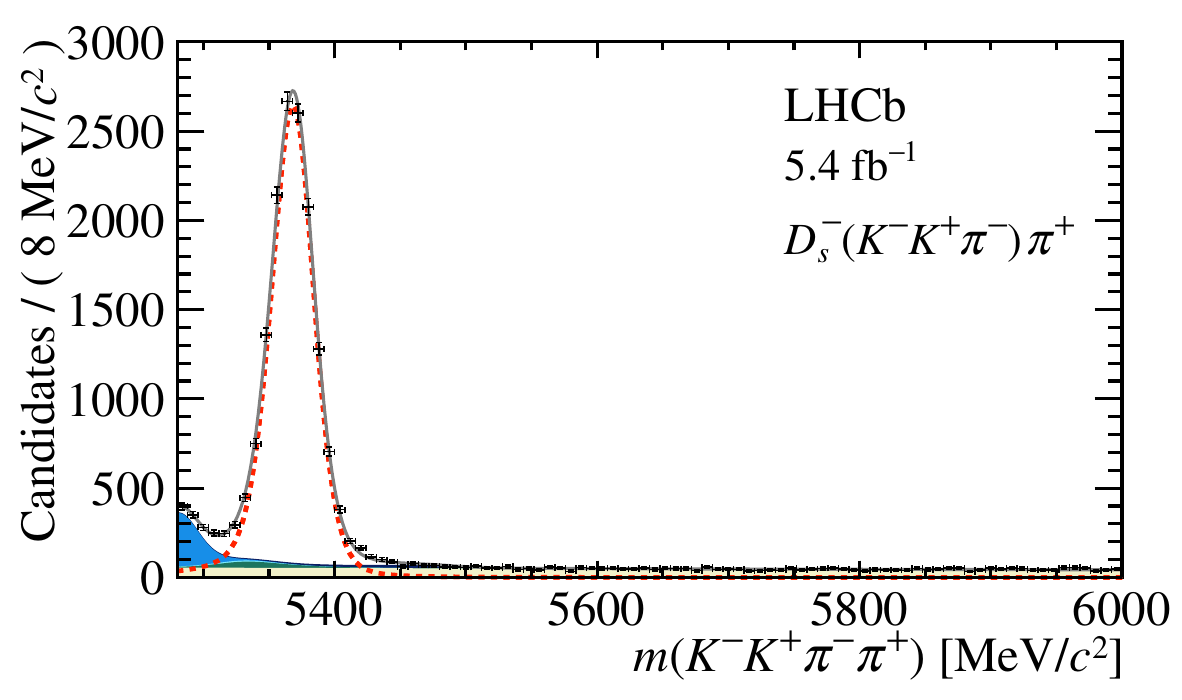}
  \includegraphics[width=.465\textwidth, clip, trim={0 0.275cm 0 0.4cm}]{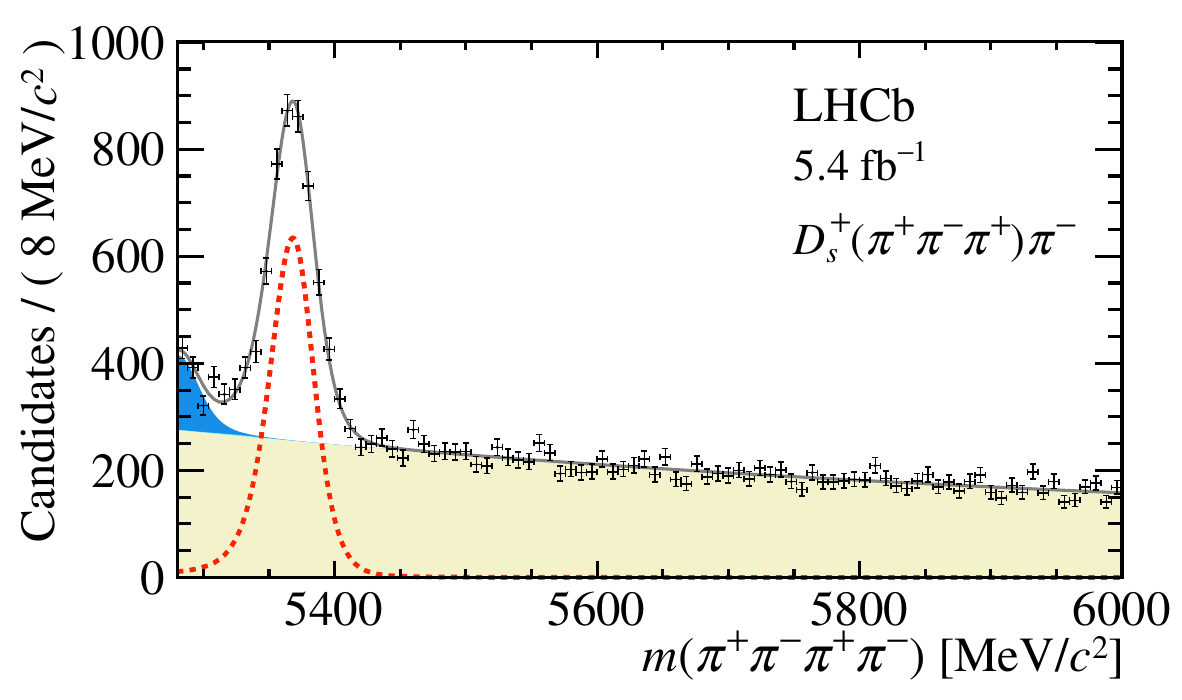}
  \includegraphics[width=.465\textwidth, clip, trim={0 0.275cm 0 0.4cm}]{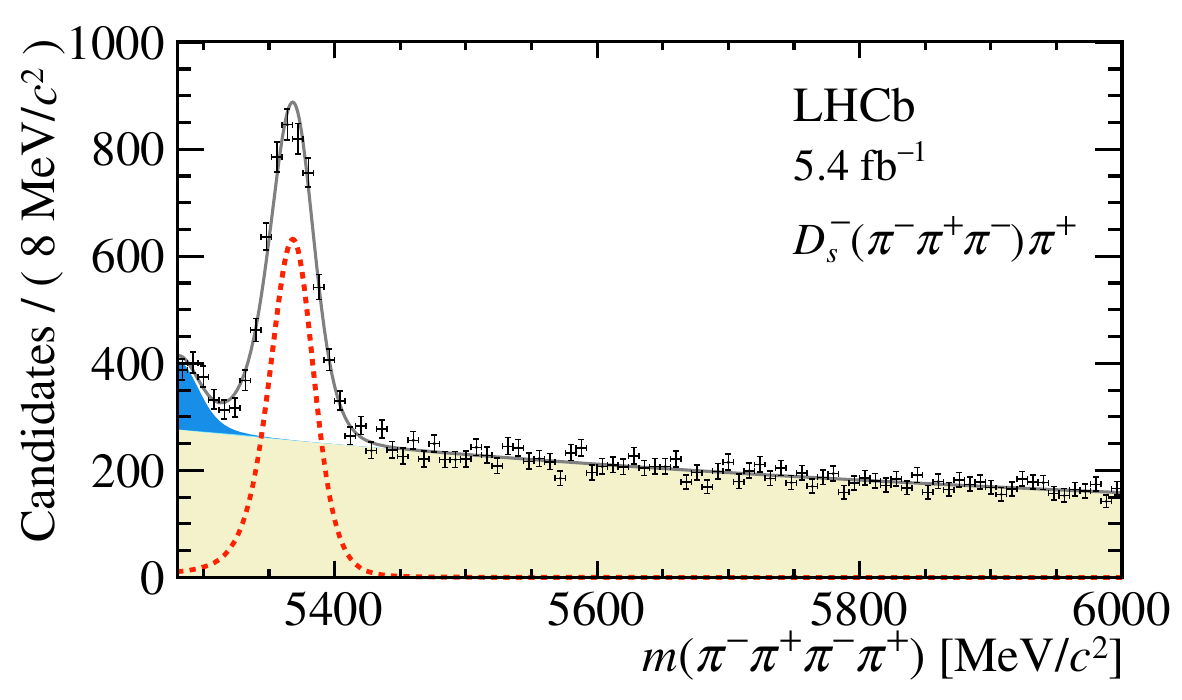}
  \caption{Invariant-mass distributions of (left) $\Dsp\pim$ and (right) $\Dsm\pip$ candidates, for the (top) $\DstoKKpi$ and (bottom) $\Dstopipipi$ decay modes, shown with the result of the invariant-mass fit.
}
  \label{fig:massfit}
\end{figure}

Distributions of invariant masses are shown in Fig.~\ref{fig:massfit}, along with the results of the mass fits.
The charge-combined signal yields are $28\,940 \pm 190$ and $7\,270 \pm 150$ for the \DstoKKpi and \Dstopipipi decay modes, respectively.
The measured value of $\araw$ is equal to $(-2.2 \pm 6.2)\times 10^{-3}$ for the \DstoKKpi decay mode and $(-0.2 \pm 1.6) \times 10^{-2}$ for the \Dstopipipi decay mode, where the uncertainties are statistical.

Following Eq.~\ref{eqn:As_untagged}, the measured $\araw$ is corrected for detection asymmetries in the $\Dsm\pip$ and $\Dsp\pim$ final states, evaluated using calibration data and simulated samples.
This is decomposed as a sum of four contributions: track reconstruction ($A_{\rm track}$), kinematic and HCAL-based selection requirements ($A_{\rm sel}$), particle-identification requirements ($A_{\rm PID}$), and \lone requirements ($A_{\rm \lone}$).
An additional contribution associated with the \hltone requirements, $A_{\rm \hltone}$, is estimated using a calibration sample of \DtoKpipi decays and found to be small.
Its effect is therefore included as a systematic uncertainty on $A_{\rm sel}$.
Table~\ref{tab:adet} summarizes the measured values of the individual contributions to $A_{\rm det}$, which are described in detail below.

For the \DstoKKpi decay mode, the track reconstruction asymmetry is obtained as the sum of three components: the single-pion, combined two-track $\Km\pip$, and single-kaon track reconstruction asymmetries.
For the \Dstopipipi decay mode, the track reconstruction asymmetries of each final-state pion are combined, assuming each pion to be uncorrelated.

The term $A_{\rm track}\left(\pim\right)$ is estimated from the asymmetry of efficiencies to fully reconstruct muons in bins of muon kinematics using samples of partially reconstructed ${\jpsi \to \mup \mun}$ decays with the tag-and-probe approach as described in Ref.~\cite{LHCB-DP-2013-002}.
One muon is reconstructed fully and the other reconstructed using limited detector information; the efficiency is estimated from the fraction of partially reconstructed muons which can be matched to fully reconstructed muons, where numerator and denominator are obtained by maximum-likelihood fits to the dimuon invariant mass.
The relative effect of hadronic interactions of pions with the detector material is assumed to have a small effect on $A_{\rm track}\left(\pim\right)$.
In contrast to pions, kaons have significant charge-asymmetric nuclear cross-sections, thus $A_{\rm track}\left(\Km\pip\right)$ is determined from the raw asymmetries of prompt \DtoKpipi and \DtoKspi decays, following the method described in Ref.~\cite{LHCb-PUB-2018-004}.
Candidates are weighted such that the kinematics of the final state particles in the \DtoKpipi (\DtoKspi) decay reflect the kinematic distributions of the signal \DstoKKpi (weighted \DtoKpipi) sample.
This kinematic weighting mitigates phase-space-dependent effects on the reconstruction efficiency.

The remaining terms are determined from charge-specific efficiencies of the corresponding requirements:
$A_{\rm sel}$ from the efficiencies of the selection requirements in simulation, and $A_{\rm PID}$ by evaluating the per-event PID efficiencies in bins of the kinematics of each particle from calibration samples of $D^{*+} \to D^0\left(\Km\pip\right)\pip$ decays, where the charge of the companion pion identifies the charge of the kaon and pion originating from the $D^0$ meson, independently of PID requirements~\cite{LHCb-DP-2018-001}.
Similarly, $A_{\rm \lone}$, arising from the charge-dependent efficiencies of the \lone trigger to select light hadrons,  is determined using reference $D^{*+} \to D^0\pip$ decays in which pions and kaons originating from the $D^0$ meson are well identified.
The trigger efficiency is computed separately for the inner and outer HCAL regions in bins of transverse energy.
As these contributions are evaluated using the same simulated samples, the corresponding statistical uncertainties are assumed to be fully correlated.
The statistical uncertainty on $A_{\rm det}$ is thereby evaluated as the quadrature sum of the largest of these statistical uncertainties with the statistical uncertainty on $A_{\rm track}$.

\begin{table}[tb]
  \centering
  \caption{Contributions to $A_{\rm det}$, averaged over magnet polarities and data-taking periods.
  The first uncertainties are statistical and the second are due to systematic sources.
  All values are given in percent.}
  \label{tab:adet} 
  \begin{tabular}{l c c }
    \toprule
     Asymmetry (\%) & $\Kp\Km\pip\pim$ & $\pip\pim\pip\pim$ \\
    \midrule
        $A_{\rm track}$ & $\phantom{-} 0.008 \pm 0.061 \pm 0.073$ & $-0.010 \pm 0.023 \pm 0.029$ \\
        $A_{\rm sel}$ & $\phantom{-} 0.031 \pm 0.063 \pm 0.080$ & $\phantom{-} 0.03\phantom{0} \pm 0.24\phantom{0} \pm 0.22\phantom{0}$ \\
        $A_{\rm PID}$  & $\phantom{-} 0.021 \pm 0.088 \pm 0.057$ & $\phantom{-} 0.08\phantom{0} \pm 0.34\phantom{0} \pm 0.17\phantom{0}$ \\
        $A_{\rm \lone}$ &  $ -0.169 \pm 0.056 \pm 0.007$ & $\phantom{-} 0.00\phantom{0} \pm 0.21\phantom{0} \pm 0.02\phantom{0}$ \\
        
        \midrule
        $A_{\rm det}$ & $-0.11\phantom{0} \pm 0.11\phantom{0} \pm 0.11\phantom{0}$ & $\phantom{-} 0.10\phantom{0} \pm 0.34\phantom{0} \pm 0.28\phantom{0}$ \\
        
    \bottomrule
  \end{tabular}
\end{table}

The systematic uncertainties on $\araw$ originate from imperfect modeling of the signal and background in the invariant-mass fits and from fixing misidentified background yields and their asymmetries.  
Each source is evaluated as the difference in $\araw$ relative to the baseline fit result after repeating the data fit accounting for each effect.
For the signal model, an alternative parametrization using the sum of a Gaussian and a modified Gaussian function from Ref.~\cite{LHCb-PAPER-2020-019} is employed.
Similarly, the low-mass background systematic uncertainty is estimated by replacing the \BdtoDspi shape with the \BstoDsstpi model.
Alternative fits are also performed allowing to vary within a Gaussian constraint all parameters fixed in the baseline fit, \ie the fraction between the two exponential functions of the combinatorial background model, the fractions defining the \BstoDsk and \BdtoDpi yields, and the misidentified background asymmetries.
Individual contributions are assumed to be uncorrelated and combined in quadrature, resulting in systematic uncertainties of $4.9\times10^{-4}$ and $1.3\times10^{-4}$ for the \DstoKKpi and \Dstopipipi decay modes, respectively.

The systematic uncertainty on the detection asymmetry is estimated for each constituent term, as listed in Table~\ref{tab:adet}.
For $A_{\rm track}$, a systematic uncertainty is assigned for the differences between muon and pion tracking efficiency asymmetries due to pion-hadron interactions according to Ref.~\cite{LHCb-PAPER-2014-053}.
A systematic uncertainty on the pion tracking efficiency derived from $J/\psi \to \mu^+ \mu^-$ decays is obtained by varying the fit model and binning scheme.
The difference in beam-crossing angles between the magnet-up and -down configurations may induce a magnet polarity- and phase-space-dependent contribution to $A_{\rm track}$, for which a systematic uncertainty is estimated.
Two different kinematic weighting schemes are used for the prompt $D^+ \to \Km\pip\pip$ and $D^+ \to \KS \pip$ samples in determining the combined two-track $\Km \pip$ asymmetry, and the resulting difference is assigned as a systematic uncertainty.
The systematic uncertainty on $A_{\rm sel}$ is evaluated by comparing its value with and without per-event weights from PID and \lone efficiencies.
A systematic uncertainty from omitting the small $A_{\rm \hltone}$ is also assigned to $A_{\rm sel}$ by evaluating the impact of HLT1 on the raw asymmetry in reference samples of prompt $D^+ \to \Km\pip\pip$ decays.
A systematic uncertainty on $A_{\rm PID}$ is assigned by varying the PID selection requirements and the binning scheme applied to the calibration samples.
The systematic uncertainty on $A_{\rm \lone}$ is entirely due to the limited size of the available reference sample.

Studies are carried out to ensure the robustness of the measurement, in which the data sample is divided and the measurement repeated in each subsample.
The data sample is divided into subsets according to the dipole magnet polarity, which is regularly changed during data-taking periods.
The data are also divided by year of data taking, in bins of the \Bs meson and companion-pion momentum and pseudorapidity, and the resonant or nonresonant regions of the \Dsm phase space.
The measurements of \asuntagged are consistent between subsamples at the scale of the statistical uncertainties in each final state.

In conclusion, the first measurement of the flavor-untagged decay-time-integrated \CP asymmetry in \BstoDspi decays is performed using $pp$ collision data collected by the \lhcb experiment at a center-of-mass energy of 13\tev, corresponding to a total integrated luminosity of 5.4\invfb.
The \BstoDspi decay is reconstructed in two \Dsm decay modes, \DstoKKpi and \Dstopipipi.
The measured asymmetries are
\begin{align*}
    \asuntaggedKKpipiCC &= \left(-1.1 \pm 6.3 \pm 1.2\right)\times 10^{-3},  \nonumber \\
    \asuntaggedpipipipiCC &= \left(-3 \pm 16 \pm 3\right)\times 10^{-3}, \nonumber
\end{align*}
where the first uncertainty is statistical and the second systematic.
The two results are combined as a mean weighted by the quadrature sums of uncertainties, yielding
\begin{equation}
    \asuntaggeddspi = \left(-1.4 \pm 5.9 \pm 1.1\right)\times 10^{-3}. \nonumber
\end{equation}

This represents the first measurement of a decay-time-integrated \CP asymmetry in hadronic \BstoDspi decays.
The measurements of \asuntagged in each final state, and their combination, are all consistent with the SM expectations and with the measured value of \assl, and exclude generic BSM contributions that induce \CP asymmetries outside of the range $-1.9$--$1.6\%$ at the $95\%$ confidence level.

\vspace{0.2cm}

\emph{Data availability}—Data associated to the plots in this publication are made available on the CERN Document Server in Ref.~\cite{LHCb-PAPER-2025-074-cds}.

\section*{Acknowledgements}
%
%
\noindent We express our gratitude to our colleagues in the CERN
accelerator departments for the excellent performance of the LHC. We
thank the technical and administrative staff at the LHCb
institutes.
We acknowledge support from CERN and from the national agencies:
ARC (Australia);
CAPES, CNPq, FAPERJ and FINEP (Brazil); 
MOST and NSFC (China); 
CNRS/IN2P3 and CEA (France);  
BMFTR, DFG and MPG (Germany);
INFN (Italy); 
NWO (Netherlands); 
MNiSW and NCN (Poland); 
MCID/IFA (Romania); 
MICIU and AEI (Spain);
SNSF and SER (Switzerland); 
NASU (Ukraine); 
STFC (United Kingdom); 
DOE NP and NSF (USA).
We acknowledge the computing resources that are provided by ARDC (Australia), 
CBPF (Brazil),
CERN, 
IHEP and LZU (China),
IN2P3 (France), 
KIT and DESY (Germany), 
INFN (Italy), 
SURF (Netherlands),
Polish WLCG (Poland),
IFIN-HH (Romania), 
PIC (Spain), CSCS (Switzerland), 
GridPP (United Kingdom),
and NSF (USA).  
We are indebted to the communities behind the multiple open-source
software packages on which we depend.
Individual groups or members have received support from
RTP (Australia), 
Key Research Program of Frontier Sciences of CAS, CAS PIFI, CAS CCEPP (China); 
Minciencias (Colombia);
EPLANET, Marie Sk\l{}odowska-Curie Actions, ERC and NextGenerationEU (European Union);
A*MIDEX, ANR, IPhU and Labex P2IO, and R\'{e}gion Auvergne-Rh\^{o}ne-Alpes (France);
Alexander-von-Humboldt Foundation (Germany);
ICSC (Italy); 
Severo Ochoa and Mar\'ia de Maeztu Units of Excellence, GVA, XuntaGal, GENCAT, InTalent-Inditex and Prog.~Atracci\'on Talento CM (Spain);
the Leverhulme Trust, the Royal Society and UKRI (United Kingdom).


\addcontentsline{toc}{section}{References}
\bibliographystyle{LHCb}
\bibliography{main}

\newpage
\centerline
{\large\bf LHCb collaboration}
\begin
{flushleft}
\small
R.~Aaij$^{38}$\lhcborcid{0000-0003-0533-1952},
M. ~Abdelfatah$^{69}$,
A.S.W.~Abdelmotteleb$^{57}$\lhcborcid{0000-0001-7905-0542},
C.~Abellan~Beteta$^{51}$\lhcborcid{0009-0009-0869-6798},
F.~Abudin\'en$^{59}$\lhcborcid{0000-0002-6737-3528},
T.~Ackernley$^{61}$\lhcborcid{0000-0002-5951-3498},
A. A. ~Adefisoye$^{69}$\lhcborcid{0000-0003-2448-1550},
B.~Adeva$^{47}$\lhcborcid{0000-0001-9756-3712},
M.~Adinolfi$^{55}$\lhcborcid{0000-0002-1326-1264},
P.~Adlarson$^{87}$\lhcborcid{0000-0001-6280-3851}
% ...

{\footnotesize \it

% ...
$^{38}$Nikhef National Institute for Subatomic Physics, Amsterdam, Netherlands\\
% ...
$^{47}$Instituto Galego de F{\'\i}sica de Altas Enerx{\'\i}as (IGFAE), Universidade de Santiago de Compostela, Santiago de Compostela, Spain\\
% ...
$^{51}$Physik-Institut, Universit{\"a}t Z{\"u}rich, Z{\"u}rich, Switzerland\\
% ...
$^{55}$H.H. Wills Physics Laboratory, University of Bristol, Bristol, United Kingdom\\
% ...
$^{57}$Department of Physics, University of Warwick, Coventry, United Kingdom\\
% ...
$^{59}$School of Physics and Astronomy, University of Edinburgh, Edinburgh, United Kingdom\\
$^{60}$School of Physics and Astronomy, University of Glasgow, Glasgow, United Kingdom\\
$^{61}$Oliver Lodge Laboratory, University of Liverpool, Liverpool, United Kingdom\\
% ...
$^{69}$Syracuse University, Syracuse, NY, United States\\
% ...
$^{87}$Department of Physics and Astronomy, Uppsala University, Uppsala, Sweden, associated to $^{60}$\\
% ...
}
\end{flushleft}




\end{document}